\documentclass[aip,rsi,reprint]{revtex4-1}
\usepackage{amsmath,amsfonts,amssymb}%
\usepackage{graphicx}%
\usepackage{natbib}
\usepackage{epstopdf}
\usepackage{placeins}
\graphicspath{{Figures/}{Figures/EPS/}}

\begin{document}

\title{Local characterization of hindered Brownian motion by using digital video microscopy and 3D particle tracking}

\author{Simon L Dettmer}
\author{Ulrich F Keyser}
\author{Stefano Pagliara}
\affiliation{Cavendish Laboratory, University of Cambridge, 19 J J Thomson Avenue, Cambridge, CB3 0HE, United Kingdom}

\date{\today}

\begin{abstract}In this article we present methods for measuring hindered Brownian motion in the confinement of complex 3D geometries using digital video microscopy. Here we discuss essential features of automated 3D particle tracking as well as diffusion data analysis. By introducing local mean squared displacement-vs-time curves, we are able to simultaneously measure the spatial dependence of diffusion coefficients, tracking accuracies and drift velocities. Such local measurements allow a more detailed and appropriate description of strongly heterogeneous systems as opposed to global measurements. Finite size effects of the tracking region on measuring mean squared displacements are also discussed. The use of these methods was crucial for the measurement of the diffusive behavior of spherical polystyrene particles (505~nm diameter) in a microfluidic chip. The particles explored an array of parallel channels with different cross sections as well as the bulk reservoirs. 
For this experiment we present the measurement of local tracking accuracies in all three axial directions as well as the diffusivity parallel to the channel axis while we observed no significant flow but purely Brownian motion. Finally, the presented algorithm is suitable also for tracking of fluorescently labeled particles and particles driven by an external force, e.g. electrokinetic or dielectrophoretic forces.
\end{abstract}


\maketitle


\section{Introduction}
In the past, the phenomenon of confined Brownian motion, sometimes also described as hindered diffusion (i.e. the decrease of diffusion coefficients in proximity to confining walls) has been studied in detail theoretically \cite{Benesch2003, Lobry1996, Bungay1973, Brenner1961, Goldman1967}, numerically \cite{Jun2006} as well as experimentally \cite{Lin2000,Sharma2010,Faucheux1994,Choi2007,Banerjee2005,Bevan2000,Eral2010,Leach2009,Ha2013}. For general cases, the theoretical treatment is quite involved and analytical predictions have been limited to very simple confining geometries such as plane walls or cylindrical channels with infinite extension. 
Past experiments have mainly focused on confirming the analytical predictions for a sphere diffusing in proximity of plane walls \cite{Lin2000,Sharma2010,Faucheux1994,Choi2007,Banerjee2005,Bevan2000}. Common to these experimental approaches is the use of video microscopy observations of single micrometer or even sub micrometer sized spherical colloids. 
Early experiments for measuring confined Brownian motion employed bright-field microscopy and were restricted to tracking in 2D and the axial position was determined from averaging over the accessible volume \cite{Faucheux1994}. Later, it became apparent that in bright-field microscopy images the axial position of spherical colloids could be extracted from comparing the radial intensity distribution of particles to prior calibration measurements on colloids stuck to a glass slide and moved axially with a piezo stage \cite{Crocker1996}. The achieved tracking accuracies using calibration measurements on 15,000 images of colloids was 150~nm axially and 10~nm laterally for 300~nm colloidal spheres. The same authors proposed the use of blinking optical tweezers to allow controlled positioning of colloids for measuring diffusion coefficients. Subsequently, this technique was applied in the study of confined Brownian motion of spherical colloids above one as well as in between two parallel glass slides \cite{Lin2000}.
More recently, the measurement of hindered diffusion close to a plane wall has been investigated at even higher precision with digital holography microscopy and total internal reflection microscopy reaching axial accuracies better than 50~nm \cite{Sharma2010, Choi2007,Banerjee2005}. For the case of free diffusion in a homogeneous medium tracking accuracies with digital holography microscopy as high as 1~nm in all three axial directions have been reported \cite{Cheong2010, Dixon2011} and hindered diffusion perpendicular to a plane wall was measured with the same accuracy using total internal reflection microscopy.
To our knowledge, so far only one study investigates the spatial dependence of diffusion coefficients in hindered diffusion of colloidal particles in a more complex geometry (in this case a closed cylinder) \cite{Eral2010}. Given the lack of video microscopy studies of hindered diffusion in the presence of more complicated boundaries, techniques for carrying out such experiments have not yet been investigated in detail. 
This is somewhat surprising as diffusion plays a major role for transport processes in the biological world that exhibits a rich variety of shapes. Mimicking these processes in vitro hence requires the realization of similarly complicated geometries. A possible practical application of knowledge of confined Brownian motion is in designing drugs in order to maximize their diffusive intake by target cells~\cite{Sugano2010}. Especially for large and hydrophobic molecules, diffusion through aqueous protein pores (facilitated transport) is a major transport mechanism~\cite{Sugano2010}. There have been sophisticated attempts to rationalize the mechanisms of channel-facilitated transport theoretically \cite{Bezrukov2000, Berezhkovskii2005} as well as experimentally \cite{Pagliara2013}. These models require knowledge of the functional form of hindered diffusion coefficients along a bulk-channel-bulk geometry for which no analytical or experimental data exists to date. The reason why it is important to discern between diffusivity values in the bulk, the channel entrance region and the channel interior is that they govern different aspects of facilitated transport. While the bulk diffusivity is responsible for transport of particles towards the channel, the diffusivity in the channel entrance region determines the particle in- and efflux to and from the channel, and finally the average translocation time is dominated by the channel interior diffusivity.\newline
Out of this consideration we feel the need to discuss methods for measuring hindered diffusion in complex geometries thus enabling future studies in this area of physical and biological importance.
The methods presented here were crucial for studying the diffusion behavior of colloids in a microfluidic bulk-channel-bulk geometry which we observed using bright-field digital video microscopy. While bright-field microscopy allows for less resolution than the before mentioned techniques that employ optical trapping and detailed scattering analysis, it is superior in the way that it is suitable for measurements on entire particle ensembles with very low computational cost, thus reducing the experimental duration. Using larger particle ensembles has the further benefit that a large volume can be observed simultaneously. \newline
In addition to the measurement of hindered Brownian motion, the presented tracking algorithm is also applicable to other phenomena such as sub-micrometer particles under electrokinetic or dielectrophoretic forces observed with fluorescence or phase contrast microscopy. Interestingly, single particle tracking can increase the sensitivity of characterization of micro-organism subpopulations by dielectrophoretic separation~\cite{Su2014}. The obtained particle displacement-vs-time curves allow for a direct extraction of the dielectrophoretic force and a simultaneous measurement of positive and negative dielectrophoresis which has advantages over techniques that use a time-dependent fluorescence intensity averaged over a large volume~\cite{Bakewell2004}. 
\newline
In this article we first describe the experimental setup used for acquiring the microscopy videos of diffusing colloids. We then present our algorithm for automated particle tracking which yields colloid trajectories in 3D. In the following, we describe how to optimally extract globally averaged colloid diffusivity and tracking accuracy from these trajectories. Here we include a discussion of finite size effects of the tracking region on measuring mean squared displacements~(\textit{MSD}s). Next, we describe the simultaneous measurement of local diffusion coefficients and tracking accuracies from the acquired trajectories for which we introduce local \textit{MSD}-vs-time curves. We then discuss how to correct the \textit{MSD}s for the influence of a stationary drift field as well as how to measure the local drift velocities. We conclude by presenting a method for measuring the system geometry \textit{in situ} and in 3D using the tracking data.

\section{Experimental setup}
\begin{figure}[htbp]
		\includegraphics[width=8.5cm]{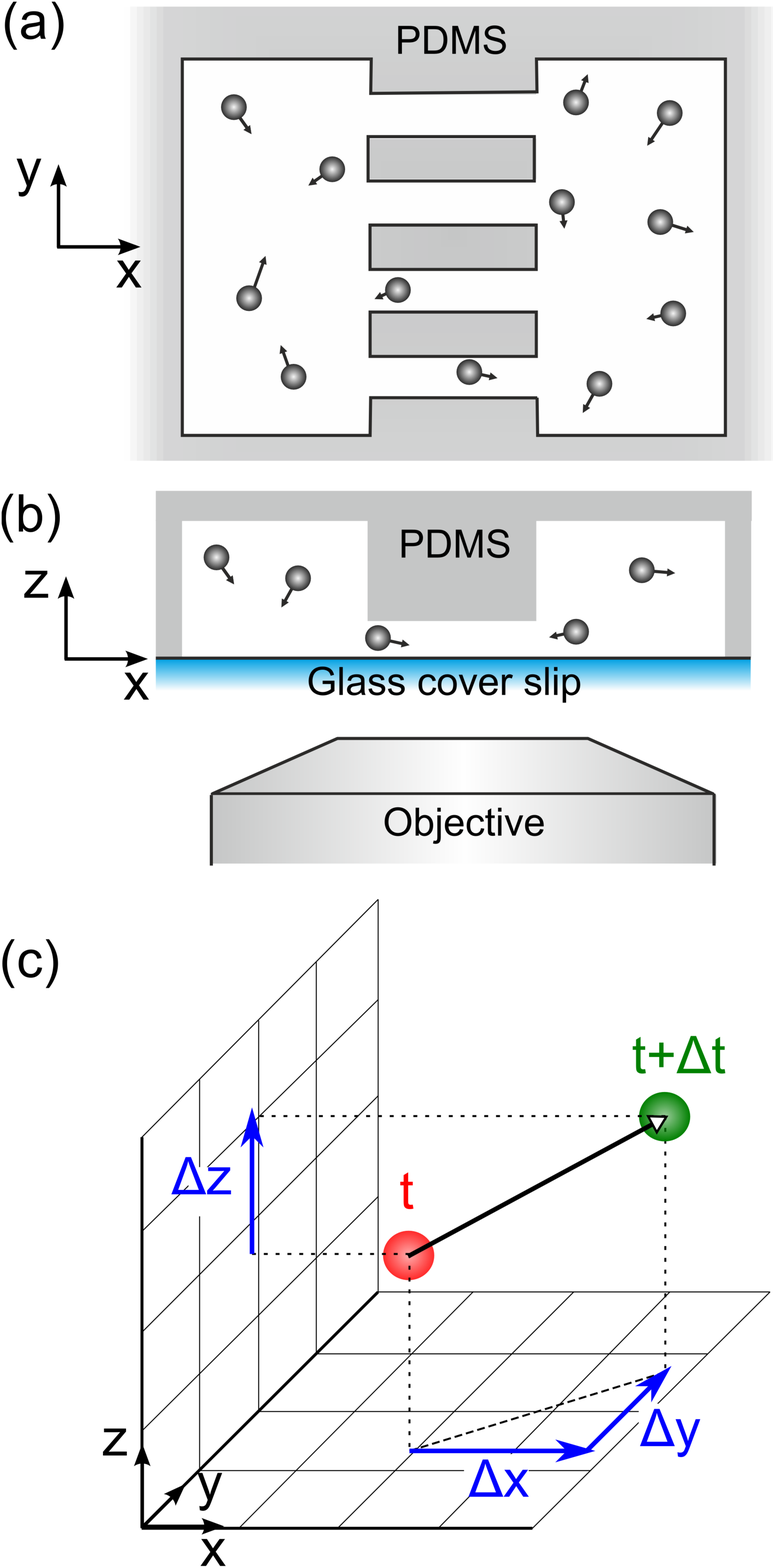}
			\caption{Cartoon illustrating the measurement of diffusivity in a microfluidic chip equipped with a complex 3D geometry. (a) The chip (viewed from top) consists of two bulk reservoirs of thickness 16~$\mu m$ separated by a 5~$\mu m$ long polydimethylsiloxane (PDMS) barrier and connected via an array of microchannels~(width $\approx 1~\mu m$). The chip is filled with a 5~mM KCl suspension of spherical polystyrene particles~(diameter = 505~nm). (b) In a side-view we see the particles freely diffusing in 3D Brownian motion in the two bulk reservoirs while being closely confined in the channels. The PDMS chip is bonded to a glass cover slip and imaged through a 100$\times$ objective. (c) Single particle tracking in x,y and z directions. We evaluate the displacements of colloids between video frames that are separated by a time lag $\Delta t$.}
	\label{fig:setup}
\end{figure}

Our microfluidic chip equipped with sub-micrometer channels was manufactured in the following way. First, an array of Platinum wires, semi-elliptical in cross section, was deposited on a Silicon substrate via focused ion beam. Each wire cross section was measured \textit{in situ} by slicing the wire at one end, tilting the sample at 63~$^\circ$C and imaging via an electron beam. Second, conventional photolithography, replica molding and Polydimethylsiloxane~(PDMS) bonding to a glass slide were carried out to define 16~$\mu m$ thick reservoirs separated by a PDMS barrier and connected by an array of channels obtained as a negative replica of the Platinum wires (Figure \ref{fig:setup}). Further details of the fabrication are reported elsewhere~\cite{Pagliara2011, Pagliara2013}. The chip was filled with spherical polystyrene particles (Polysciences (Warrington, PA), $(505\pm8)$~nm diameter) dispersed in a 5~mM KCl solution and continuously imaged through an oil immersion objective (100$\times$, 1.4 N.A., UPLSAPO, Olympus). Illumination was provided from above by an LED light (Thorlabs MWLED). The transmitted light was collected by the objective and coupled to a CCD camera (frame rate of 30~fps and magnification of  16.7~px/$\mu$m in $x-$ and 14.7~px/$\mu$m in $y$-direction after video compression). Experiments were automated by using a custom-made program based on LabVIEW for positioning and video acquisition as reported elsewhere~\cite{Pagliara2013} and performed overnight in order to reduce the noise level.

\section{3D Particle Tracking}
\begin{figure*}[htbp]
\includegraphics[width=17cm]{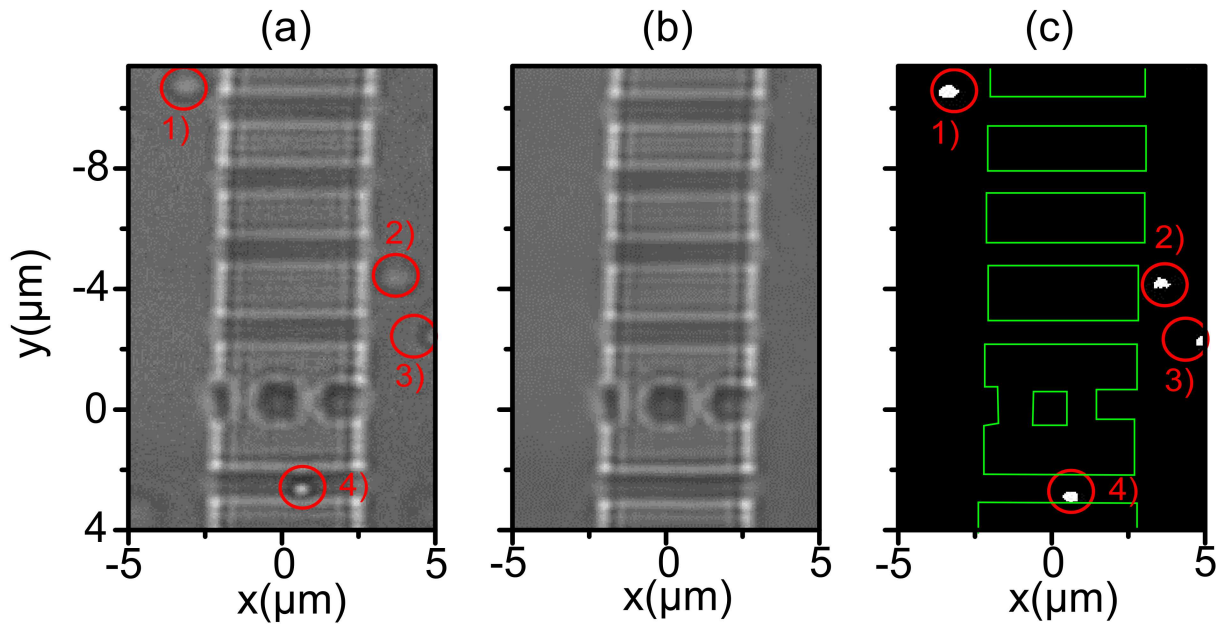}
\caption{Image processing and particle tracking presented for a typical video frame. (a) shows a single frame from a video acquired with bright field microscopy. (b) Averaging the light intensity over 100 frames spread throughout the video yields an appropriate background image. (c) Subtracting the background image and applying a threshold first and then two 3$\times$3 pixel erosions allows to identify the individual particles and their positions. To determine the exact particle center positions, in a first step the center of mass of the white discs in (c) is calculated and then this estimate is refined by averaging the light intensity from the image (a) in a 3$\times$3 pixel$^2$ box around this estimated position. Particles are marked by the numbered red circles in (a) and (c). The contours of the channels are superimposed on the transformed image by green lines for illustration.}
	\label{fig:img-transform}
\end{figure*}

In order to achieve automated 3D particle tracking we developed a custom-written LabVIEW routine that works in three steps. In the first step the particles are distinguished from the background and exactly located in the plane. Using these particle positions as an input an independent algorithm determines the axial position. In the third and last step the 3D positions of particles in subsequent video frames are linked into trajectories.
For homogeneous backgrounds the method of choice is a cross-correlation and centroid algorithm. This approach has been widely employed for particle tracking with different microscopy techniques \cite{Carter2005}. In the case of strongly inhomogeneous backgrounds, however, the cross-correlation algorithm performs very badly. Inhomogeneous backgrounds can be due to microfluidic structures in the observed region, for instance the PDMS barrier with the array of microchannels (Figure~\ref{fig:img-transform}(a)). In most of the previous studies, researchers focused on measuring hindered diffusion of colloids either above a plane wall or restricted to the interior of very long channels thus leading to a homogeneous background. However, for studying the diffusion behavior at the bulk-channel interface it is necessary to track particles in the channels and bulk simultaneously. For this we have to deal with inhomogeneous backgrounds at the structure edges. There are mainly two reasons for the bad performance of cross-correlation algorithms in inhomogeneous backgrounds. Firstly, sharp contrasts at edges of background structures in the image tend to get high cross-correlation values. This would necessitate applying a high threshold to the image in order to prevent these edges to be mistaken for actual particles. Hereby the number of detected particles in the regions distant from the edges is dramatically reduced. Furthermore, the average light intensities vary across different regions of the image.  As the same threshold is applied to the entire image, particles in darker regions are less likely to be detected or might not be detected at all.
To circumvent these difficulties we subtract a background image and directly apply a threshold and centroid algorithm without computing the cross-correlation (Figures \ref{fig:img-transform} (a)-(c)). For removing background noise we apply four 3$\times$3 pixel erosions using the LabVIEW IMAQ Remove Particle Routine. The threshold and number of erosions is chosen by manually inspecting the effect on different test images in the beginning of the experiment analysis. Subtracting the background proves essential as it allows the choice of a comparatively low threshold and removes additional noise from the edges of the background wall structure.
Ideally, the background image should not contain any colloids as using such a background would impede the tracking of particles in regions where the background contains colloids close to the focal plane. Practically, it proves difficult to directly acquire a colloid free background image as we would need to fill the chip after it has been positioned under the microscope or otherwise exactly reposition it after filling. Instead, a clever way to obtain an appropriate background image from a microscopy video is to calculate it by averaging the light-intensities over the entire course of the video. In our case we use 100 images separated equally throughout the video (Figure \ref{fig:img-transform}(b)). When tracking particles in fluorescence microscopy videos, however, there is no benefit in subtracting a background image since structures other than the particles themselves will not fluoresce. Instead, noise in the image can be reduced by convoluting the image with a Gaussian kernel which corresponds to calculating the cross-correlation with the Gaussian point-spread function of fluorescent beads~\cite{Smal2010}.

\begin{figure}[htbp]
		\includegraphics[width=8.5cm]{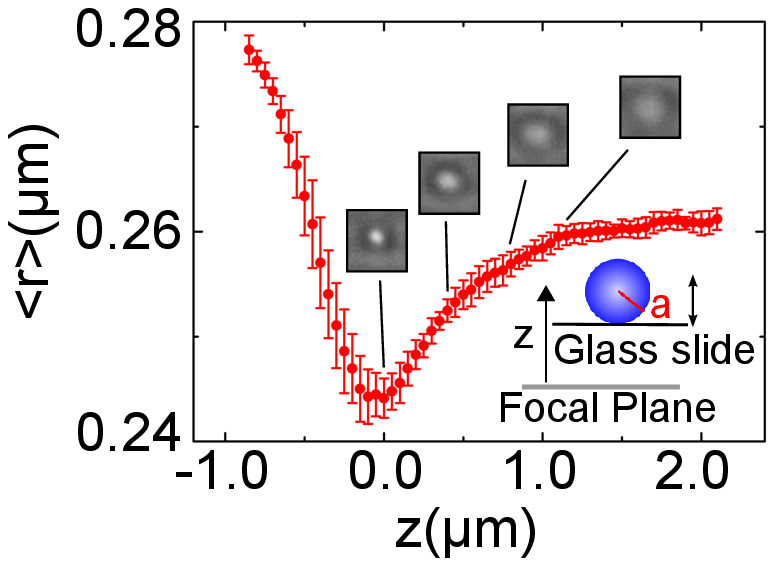}
	\caption{Calibration curve for determining axial positions for 3D particle tracking. Particles stuck to the glass slide are moved axially through the focal plane in 50~nm steps with help of a piezo stage. This is depicted by the illustration in the bottom right corner. The intensity averaged radius $\langle r \rangle$ of the colloids increases almost monotonously for $z$-positions above the focal plane. The few outliers at larger distances from the focal plane where the averaged radius decreased were excluded thus enforcing a monotonous curve. Piecewise linear fitting of the measured average radius to this calibration curve then allows continuous axial tracking. The insets show the raw microscopy images at different heights of one of the colloids used for calibration.}
	\label{fig:average_radius}
\end{figure}

For determining the axial position from the 2D images we follow the approach of Crocker and Grier \cite{Crocker1996} and consider the moments of the radial light-intensity distribution around the colloid center. To calibrate this measurement we record images of several colloids stuck to a glass slide and use a piezo stage to move them axially through the focal plane. Rather than considering the data-intensive 2D-distribution of two separate moments, we find it sufficient to use the first moment (intensity averaged particle radius) alone, achieving accuracies comparable to those reported in the original work \cite{Crocker1996}. The calibration curve (Figure \ref{fig:average_radius}) shows a decreasing and an increasing part separated by a single local minimum. As the only distinguished axial position of a colloid is in the focal plane, we define the minimum of this curve as $z=0$. In experiments the glass slide is brought into the focal plane (Figure~\ref{fig:setup}) so that negative axial positions are physically prohibited and thus only the increasing part on the right-hand side of the calibration curve is accessible to the particles. Restriction to this increasing part enables us to obtain continuous $z$-positions from piecewise linear fitting to the calibration curve. \newline
After the particles are identified and located successfully, their positions in successive frames are linked into trajectories by placing boxes around the identified particle centers in each frame. Candidates for subsequent particle positions in the next frame are only searched for within the box. If no candidate particle can be identified the trajectory is terminated. In case more than one particle is present in the box the one closest to the box center is selected. The likelihood of this event depends on the box size and particle concentrations used. In our experiment we estimate that it is not uncommon and occurs in $\sim 32\%$ of cases (see Supplemental Information~\cite{Supplemental}).  The position of the box is updated after each frame by centering it on the last particle position and it is chosen large enough so that the particles do not exit it between two consecutive frames. Typical step sizes in between frames were on the order of $200~nm\approx 3~px$. An adequate box size (in our case 30$\times$30~px$^2$) can be chosen both by manual inspection as well as a theoretical estimate on the expected particle displacements employing the theoretically expected distribution of displacements from bulk diffusion. Our theoretical analysis of the used linking algorithm shows that it is very effective and works well even at very high particle concentrations with linking mistakes occurring only in extremely rare cases (see Supplemental Information~\cite{Supplemental}). In cases where the motion is not predominantly Brownian but rather has a large drift contribution, such as in electrokinetic flows, the linking algorithm could be modified slightly. In these cases it would be reasonable to center the box not on the previous particle position but rather on an assumed position in the next frame, extrapolated from the measured displacement between the last two frames under the assumption of constant velocity.\newline
Finally, combining all the mentioned steps, this algorithm provides us with 3D particle trajectories and time steps discretized by the time lag $\tau$ between consecutive video frames: $\vec{r}(t), (t=0,\tau,2\tau,3\tau,N\tau$)  where \textit{N}+1 denotes the number of data points recorded for the individual trajectories and $\tau=33~ms$ in our experiments.

\section{Measuring globally averaged diffusivity and tracking accuracy}
Let us first discuss the measurement of globally averaged quantities describing Brownian motion. For this purpose we will consider complete trajectories of individual particles.
To characterize Brownian motion, the mean squared displacement (\textit{MSD}) is the most widely used quantity. In one dimension it is given by

\begin{equation}
MSD_i(t)\equiv \left \langle \Delta r_i^{2}(t)   \right \rangle = \left \langle \left( r_i(t)-r_i(0) \right)^2 \right \rangle  ; i\in \left \{ x,y,z \right \}.
\label{eq:1}
\end{equation}

In the limit of an infinite fluid, constant diffusion coefficient $D_i$ and assuming the absence of any net drift (the effect of drift will be discussed in section~\ref{sec:Local_Measurements}), the \textit{MSD} is proportional to time, with the proportionality constant $2D_i$. In practice, this linear behavior is a good approximation if the space that is explored by the particle is small compared to the extension of the fluid and the scales on which the diffusion coefficient varies. In our experiment the latter assumption would be fulfilled for particles exploring either only the bulk reservoirs or only the channels. In the statistical analysis these events need to be treated separately. For the time scales for which we could track particles, the explored space in $x$-direction was always small compared to our field of view and thus the former assumption justified. For movement in $y$- and $z$-direction, however, the presence of boundaries (channel walls and glass slide) leads to a saturation of the \textit{MSD}s at large times.
While for these short times the proportional relationship holds true for the actual particle positions, it needs to be modified for the measured positions in order to include systematic error contributions originating from finite tracking accuracy and image acquisition time~\cite{Savin2005}. Denoting the finite image acquisition time with $\theta$ and the variances on the position measurements with $\sigma_i^2$, the  corrected relation is given by

\begin{equation}
MSD_i(t)=2D_it +\left \{2\sigma_i^2-\frac{2}{3}D_i\theta  \right\}.
\label{eq:2}
\end{equation}

The diffusion coefficients $D_i$ and localization uncertainties $\sigma_i$ for each trajectory can then be obtained from a linear fit to the \textit{MSD}-vs-$t$ curves. This yields a distribution of diffusion coefficients and tracking accuracies around the true values due to the limited statistical accuracy caused by finite trajectory lengths. Assuming that all particles diffuse at the same speed (i.e. they are monodisperse) the true values can be determined from the mean and its standard error. For polydisperse particles the different diffusion coefficients can be evaluated from Gaussian fits to the separate peaks of the distribution of diffusion coefficients corresponding to the different particle sizes. The optimal number of \textit{MSD} points to be included in the linear fit depends on the ratio of the static \textit{MSD}-contribution ($\left \{2\sigma_i^2-\frac{2}{3}D_i\theta  \right\}$) to the dynamic one ($2D_it$). This ratio is known as the reduced error \cite{Michalet2010}. In our case the reduced error was smaller than $0.6$ in the bulk and even smaller than $0.06$ in the channels, suggesting that fitting to only the first two points of the \textit{MSD} curve is optimal. The reason for the existence of an optimal number of fitting points is as follows. On the one hand the statistical accuracy of the \textit{MSD} values decreases with increasing time because less data points become available for averaging. On the other hand if the reduced error is large, the slope arising from the diffusive motion will be comparatively small and longer time intervals are needed to determine it accurately. For example for a reduced error equal to three, it would be optimal to use the first six points of the \textit{MSD} curve. While of course \textit{a priori} this reduced error is not known exactly, an educated estimate of tracking accuracy and diffusion coefficient is often sufficient to determine an adequate number of fitting points. Following the analysis, the reduced error can be recalculated and if necessary the analysis repeated with a different number of fitting points~\cite{Michalet2010}. For theoretical expressions for the reduced errors we refer to the work of Michalet~\cite{Michalet2012}.

\subsection{Determining the \textit{MSD}-vs-time curves}
Calculating the \textit{MSD} values for a trajectory consisting of \textit{N}+1 recorded positions is normally done by taking into account all available displacements that are separated by the desired time lag \cite{Michalet2010}:

\begin{equation}
MSD_i(n\tau)=\frac{1}{N+1-n}\sum_{k=0}^{N-n}\left [ r_i((k+n)\tau)-r_i(k\tau) \right ]^2.
\label{eq:MSD_sampling}
\end{equation}

For estimating the errors on the measured \textit{MSD} curves we use the theoretical variation that arises from the stochastic nature of the process. For this purpose we define the expected relative error \textit{q}:

\begin{equation}
q_{1D}(n)=\frac{\sqrt{Var(MSD_i(n\tau))}}{\langle MSD_i(n\tau) \rangle}.
\label{eq:4}
\end{equation}

\begin{figure}[htbp]
		\includegraphics[width=8.5cm]{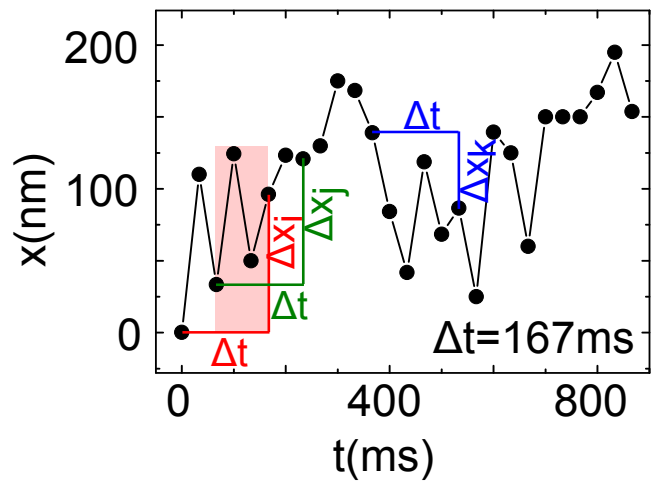}
	\caption{Scheme of the measurement of \textit{MSD}s for individual trajectories according to Equation~(\ref{eq:MSD_sampling}). The position-vs-time data points of a complete particle trajectory are analyzed in the following way: for a fixed time lag, all displacements in the trajectory, at different time intervals, are averaged to give the \textit{MSD} value for that time lag. Some time intervals used in the sampling will have overlapping parts~(marked by the red box) and thus contain the same data points. This leads to statistical dependencies between the different displacements sampled and thus a reduced information content of these samples as compared to statistically independent ones. We need to consider this effect when calculating the expected error on the \textit{MSD} values which will be larger than the $\sim1/\sqrt{N}$ standard error of the mean of uncorrelated samples.}
	\label{fig:statistical_dependence}
\end{figure}

To calculate \textit{q}, it is necessary to take into account the statistical dependence between different terms in the sum of Equation (\ref{eq:MSD_sampling}). These dependencies arise because the different terms contain overlapping parts of the trajectory (red box in Figure \ref{fig:statistical_dependence}). For the special case of isotropic diffusion and 2D-\textit{MSD}s this calculation was done by Qian et al. \cite{Qian1991}. From this we can easily deduce the expected error for the 1D case. Assuming isotropic diffusion, the 2D \textit{MSD}s can be expressed as a simple sum of 1D \textit{MSDs}:

\begin{align} 
MSD_{2D}(n\tau)&=\langle \Delta x^2(n\tau) + \Delta y^2(n\tau)  \rangle \nonumber \\&= MSD_x(n\tau)+MSD_y(n\tau) \nonumber \\&=2MSD_{1D}(n\tau).
\label{eq:5}
\end{align}

This directly gives us the relationship

\begin{equation}
q_{1D}=\sqrt{2}q_{2D}.
\label{eq:6}
\end{equation}

Into Equation (\ref{eq:6}) we can insert the expression for $q_{2D}$ derived by Qian et al. \cite{Qian1991}, giving:

\begin{equation}
q_{n,1D}=\sqrt{2}\sqrt{\frac{2n^2+1}{3n(N+1-n)}}.
\label{eq:7}
\end{equation}

\section{Finite size effects on \textit{MSD}s}
\begin{figure}[htbp]
	\centering
		\includegraphics[width=8.4cm]{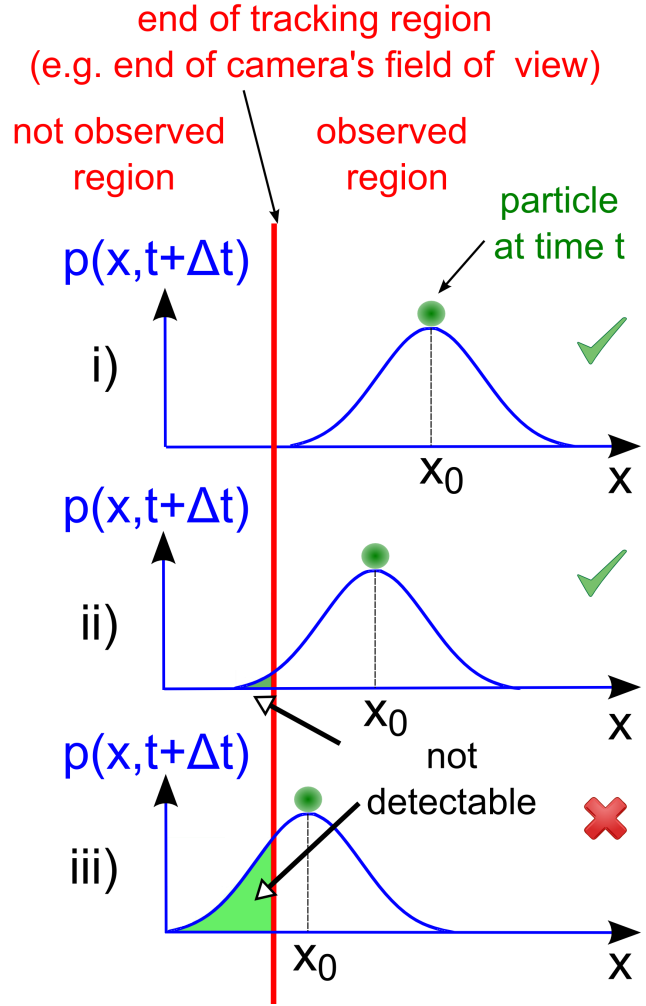}
	\caption{Illustration of finite size effects on \textit{MSD}s. Generally, the region in which particles are observed is finite, e.g. limited by the finite CCD chip size. The finite size of the tracking region can bias the measurement of \textit{MSD}s towards smaller values. Shown here are three scenarios that differ in the distance of the particle from the boundary of this region (red line) at time $t$. The probability density for finding a particle at a position $x$ at time $t+\Delta t$ is plotted in blue. Depending on the distance from the boundary there is a finite probability that the particle will exit the observable region and thus be lost from tracking at time $t+\Delta t$  (the green area under the curve). If this happens, the displacement of the particle will not enter the sample set for averaging the \textit{MSD} value for $\Delta t$. In this way the true displacement distribution is artificially truncated by the finite size of the tracking region. To avoid a sampling bias, recorded displacements of particles are only sampled for the \textit{MSD} values if the probability to exit the observable region is below a chosen threshold (we used $5~\%$). Practically, this means that the particle needs to be sufficiently far from the boundary (cases i) and ii)). On the other hand, if the particle is very close to the boundary (case iii)), there is a significant probability that it will exit the tracking region and the real distribution will be severly truncated. Therefore, we discard displacements starting from these positions.}
	\label{fig:boundary_bias}
\end{figure}

While for (infinitely) large tracking regions the \textit{MSD} values are best determined using Equation~(\ref{eq:MSD_sampling}), naively counting all observed displacements will lead to a bias of the \textit{MSD}s towards lower values in the case of a limited tracking region. The reason for this effect is that when a particle is located close to the boundary of the tracking region, the observable displacements towards the boundary are truncated at the distance of the particle center from the boundary. Particles experiencing larger jumps will exit the tracking region and thus be lost from tracking (Figure \ref{fig:boundary_bias}). The boundary of the tracking region we consider here can be either the boundary of the field of view in the plane or the limited working depth in axial tracking. The latter is typically a lot smaller than the former and thus the problem occurs more severely in axial direction. \newline
In order to solve the problem of the biased \textit{MSD} measurements we introduce an excluded volume and only sample displacements of particles located in a region where the initial particle position is sufficiently far from the boundary so that the probability to exit the tracking region becomes negligible. For quantifying sufficiently far and negligible, we can use the propagator of free diffusion. Provided with the diffusion coefficient $D_i$ it is straightforward to give a threshold on the displacements that will not be exceeded with any desired certainty $(1-\alpha)$. The easiest option is to choose the bulk diffusion coefficient $D_0=k_BT/6\pi\eta a$. For the confidence level we take the widely used convention for statistical significance, $\alpha=5~\%$, which is comparable to the statistical fluctuations of the \textit{MSD} measurements. For the bulk diffusivity we get $D_0=0.8~\mu m^2/s$. Using these values gives a distance of $450~nm$ per frame.\newline To summarise, we avoid sampling displacements of particles close the tracking region boundary because of a sampling bias that occurs through the loss of tracking at the edges. 
 
Naturally, the number of data points for evaluating the \textit{MSD} values will decrease with larger excluded volumes at the boundaries and therefore decrease the statistical accuracy. It is therefore of interest to find an optimized excluded volume at the boundaries that prevents measuring biased values but does not exclude too many data points from the analysis. This can be achieved iteratively by starting the analysis with the excluded region derived from the bulk diffusion coefficient and using the computed $D$ for updating the excluded region for the next iteration of the analysis.

\section{Measuring local diffusivity and tracking accuracy}
\label{sec:Local_Measurements}
The global measurements described above give an accurate value of the average diffusivity and tracking accuracy in the system. However, for strongly heterogeneous systems, like our complex 3D geometry or geometries involving largely different dimensions, such an average is not very informative. It gives only a coarse-grained view of the system. Therefore, we measured diffusivity and tracking accuracy locally.\newline
The main idea to achieve this is to assign measured particle displacements to positions as done e.g. in the work of Eral et al.~\cite{Eral2010}. In order to do this we need to bin the coordinates in our system. Subsequently, we take all recorded particle trajectories, choose a fixed time lag ($t=n\tau,n\in \mathbb{N}$). Then we assign every displacement $\Delta r$ separated by this time lag to the bin closest to center between the consecutive particle positions. Averaging over the displacements for each time lag yields localized mean squared displacements:
\begin{equation}
MSD_i(\vec{r},t)\equiv \langle \Delta r_i^2 \rangle (\vec{r},t).
\label{eq:local_MSD}
\end{equation}

Allowing for the spatial variance of diffusion coefficients and tracking accuracies the time-dependence of the \textit{MSD} values, Equation~(\ref{eq:2}) generalizes to:

\begin{equation}
MSD_i(\vec{r},t)=2D_i(\vec{r})t
+\left \{2\sigma_i^2(\vec{r})-\frac{2}{3}D_i(\vec{r})\theta \right \}.
\label{eq:8}
\end{equation}

\begin{figure}[htbp]
	\centering
		\includegraphics[width=8.5cm]{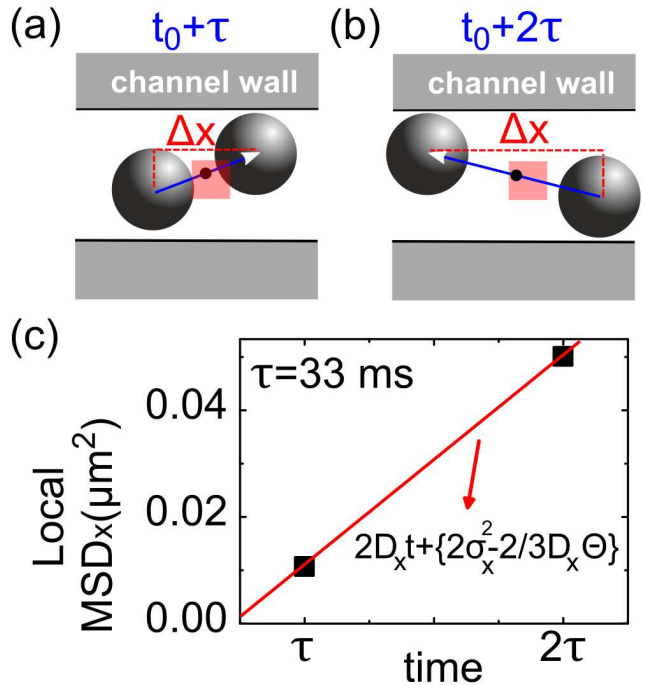}
	\caption{Illustration of the method for measuring local diffusion coefficients. Many different particles are tracked over a long period of time and the recorded displacements after time lags of one video frame (a) and two frames (b) are assigned to the bin into which the midpoint of the displacement vector falls (marked by the box). For clarity the displacement analysis is only shown for diffusion along the channel axis ($x$-direction). (c) This procedure yields the first two points of the local \textit{MSD}-vs-t-curve for each bin. By performing a linear fit to these two points we obtain the diffusion coefficient from the slope and the tracking accuracy from the offset of the line.}
	\label{fig:local_MSD_principle}
\end{figure}

From this relationship we can deduce the local diffusion coefficients and tracking accuracies by a linear fit as for the individual trajectory measurements (Figure~\ref{fig:local_MSD_principle}). As with the tracking accuracy $\sigma_i(\vec{r})$ and diffusion coefficient $D_i(\vec{r})$ we have two spatially varying unknowns in Equation~(\ref{eq:8}), the second point of the \textit{MSD} curve is required for evaluating both tracking accuracy and diffusion coefficient. This is opposed to the common practice of directly relating the first point $MSD(\vec{r},\tau)$ to the diffusion coefficient by either completely neglecting the systematic error contribution in Equation~(\ref{eq:8}) or at least using an averaged tracking error $\sigma_i$ determined by global measurements or tracking of immobilized particles. This is only justifiable for high tracking accuracies and short exposure times. 

In contrast to the statistical dependence found in global measurements~(Figure~\ref{fig:statistical_dependence}), different displacements contributing to the local \textit{MSD} values will be uncorrelated as a particle will in general not stay in a single bin for more than one video frame. For this reason the error on MSD values can be determined directly from the standard error of the mean and simple error propagation leads to the errors for the quantities derived from the linear fit.
By using Equation~(\ref{eq:8}) we implicitly assume that the length scales over which displacements are collected for a certain bin is small compared to the length scales on which the diffusion coefficients and tracking accuracies vary. However, this does not pose a fundamental problem as the variations on smaller length scales will simply be averaged out but become measurable when using a higher video frame rate. The spatial resolution with which the diffusion coefficients and tracking accuracies can be measured can easily be deduced from the data. It is determined by the distance that a particle travels within the observed time lag:
\begin{equation}
Res_i(\vec{r})\approx\sqrt{MSD_i(\vec{r},t)}.
\label{eq:spatial_resolution}
\end{equation}
This is because the traveled distance is a measure for the volume over which displacements are sampled for the different bins. Naturally, the resolution must be limited by the finite extension of the sampling volume.
Simultaneously determining local diffusion coefficients and tracking accuracies requires the first two data points so $t=2\tau$. In case the tracking accuracy found by the linear fit to the \textit{MSD}s does prove to be homogeneous, the resolution can be further increased by inserting the average tracking accuracy into Equation~(\ref{eq:8}) and then only using the first \textit{MSD} point to determine the diffusion coefficients, i.e.  choosing $t=\tau$.

\begin{figure*}[!htb]
		\includegraphics[width=14cm]{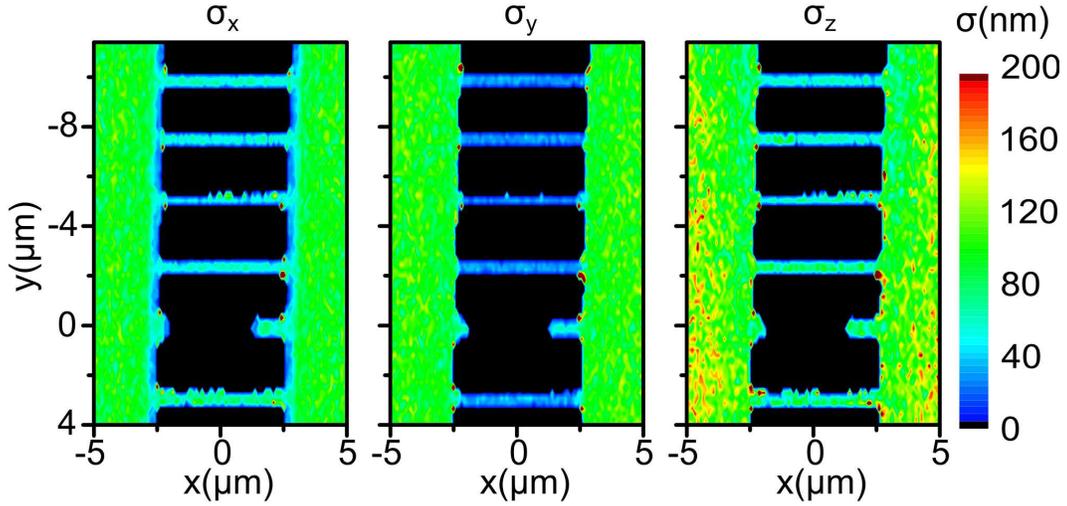}
	\caption{Color maps of local tracking accuracies. The color maps show lateral ($\sigma_x$ and $\sigma_y$) and axial ($\sigma_z$) accuracies on the position measurement which where determined from the local \textit{MSD}-vs-time curves. The accuracy decreases in the bulk reservoirs as a larger axial region is accessible to the particles and accurate tracking becomes more difficult with increasing distance from the focal plane. The best accuracy is achieved within the channels because the particles are confined close to the focal plane, giving tracking accuracies of $\sim$60~nm in $x$-, $\sim$40~nm in $y$- and $\sim$60~nm in $z$-direction. The apparent roughness of the third channel from the top is due to the statistical uncertainty arising from the small number of recorded particle positions (compare Figure~\ref{fig:countmaps}(b)).}
	\label{fig:tracking_accuracies}
\end{figure*}

\begin{figure}[!hb]
	\centering
		\includegraphics[width=7cm]{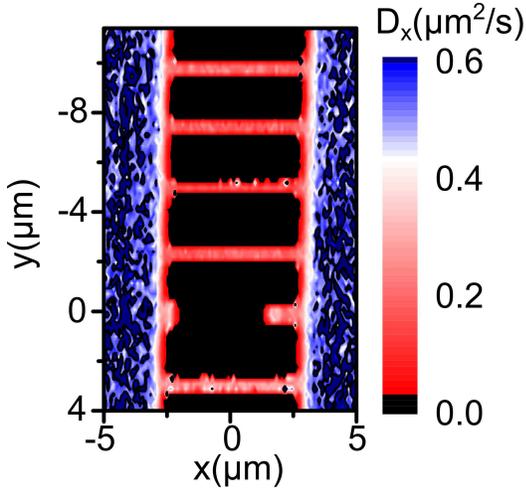}
	\caption{Color map of the local diffusion coefficients parallel to the channel axis ($D_x$). The local diffusion coefficients were determined from a linear fit to the local \textit{MSD}-vs-time curves. The diffusivity inside the channels is significantly reduced as compared to the value in the bulk reservoirs. The low (red) diffusivity in the channels and high (blue) diffusivity in the bulk are separated by a transition region of intermediate (white) diffusivity that runs parallel to the PDMS barrier walls. The apparent roughness of the third channel from the top is due to the statistical uncertainty arising from the small number of recorded particle positions (compare Figure~\ref{fig:countmaps}(b)).}
	\label{fig:Dx_colormap}
\end{figure}

For our experiment, the measured tracking accuracies are presented in Figure~\ref{fig:tracking_accuracies} and the diffusion coefficients in $x$-direction in Figure~\ref{fig:Dx_colormap}.  The tracking accuracy indeed proved to be non-uniform with the tracking error being twice as large in the bulk reservoirs as compared to the channels ($\sim~100~nm$ vs. $\sim~50~nm$, see Figure~\ref{fig:tracking_accuracies}).  While we can report the diffusivity $D_x$ parallel to the channel axis, the diffusivity in perpendicular direction ($D_y$ or $D_z$) could not be measured at the used low frame rate. The reason for this is that after a time lag of two frames (${2\tau=66~ms}$), the average traveled distance of the particles was on the order of the channel width and thus the perpendicular \textit{MSD}s (in $y$- or $z$-direction) must therefore be expected to saturate rather than showing a linear behavior. This does not significantly effect the certainty in establishing the perpendicular tracking accuracy, however, because its estimation relies on the \textit{MSD}-vs-t behavior on timescales shorter than the frame distance ($t\leq \tau$). In any case, influences of the channel confinement on the extracted tracking accuracies can only lead to an overestimation of their values due to a possible underestimation of the linear contribution from diffusive motion to the \textit{MSD}s at shorter timescales.
For the diffusion parallel to the channel axis ($D_x$) we find that the diffusion coefficients are significantly reduced inside the channels ($0.2~\mu m^2/s$, red in Figure~\ref{fig:Dx_colormap}) as compared to the value found in the bulk reservoirs ($0.5~\mu m^2/s$, blue in Figure~\ref{fig:Dx_colormap}). These two different areas are separated by a transition region of intermediate diffusivity (white in Figure~\ref{fig:Dx_colormap}). 
The data shows that neglecting the finite tracking accuracy would have been permissible only inside the channels due to the small reduced error in that region. In the bulk region, naively using the single step \textit{MSD} values and neglecting the tracking accuracy would have lead to a measured diffusivity of $0.7~\mu m^2/s$ rather than the actual value of $0.5~\mu m^2/s$. Therefore, a correct simultaneous treatment of these two regions is not possible with only the first \textit{MSD} point by either neglecting the static error or using a global one. This of course means that the interesting channel-bulk interface could not have been characterized correctly without using the second \textit{MSD} point.
Analysis of the raw \textit{MSD} values according to Equation~(\ref{eq:spatial_resolution}) revealed the achieved sampling volume and thus the spatial resolution were approximately 2~pixel$\approx$130~nm. This value should not be confused with the tracking accuracy that was around 1~px.

\section{Correcting the MSD for drift and determining local drift velocities}

\begin{figure}[p]
	\centering
		\includegraphics[width=7cm]{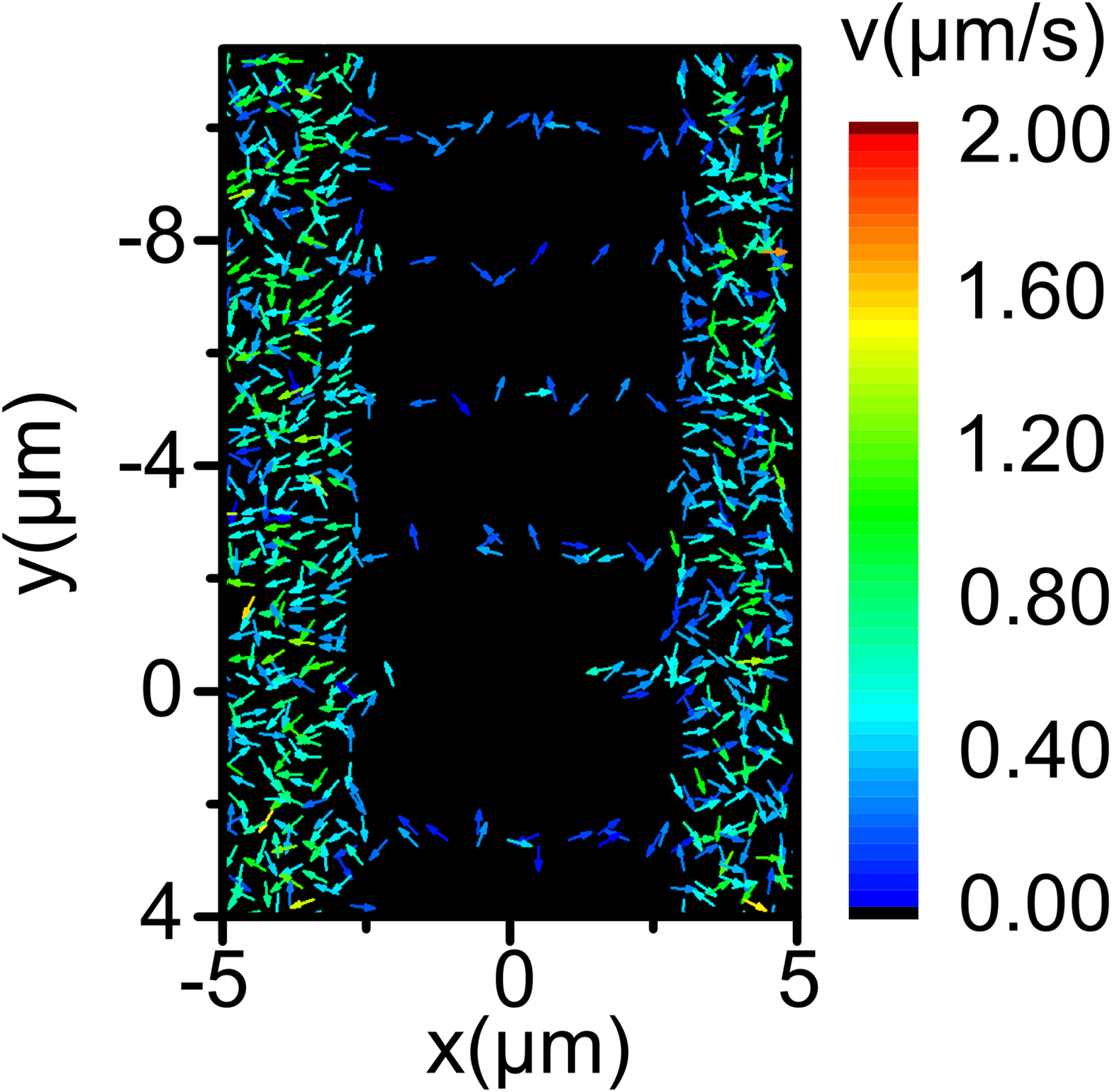}
	\caption{Mapping of local drift velocities. The local drift velocities were calculated from the local mean displacements (\textit{MD}s) according to Equation~(\ref{eq:Driftvelocity}). The arrows indicate the direction of the local drift velocity and the flow magnitude is encoded by the color scale. Within measurement accuracy there was no significant drift which is shown by both the small measured drift magnitude and the disorder of the arrows. Comparing the \textit{MSD}s associated with these drift velocities with the ones associated with the undirected Brownian motion, we find that the total \textit{MSD} and thus the particle movement is dominated by purely Brownian motion.}
	\label{fig:flow_field_color_mapped}
\end{figure}

So far we have assumed the absence of any drift. Additionally allowing for a net drift of a stationary velocity $\vec{v}(\vec{r})$, there is a contribution to the \textit{MSD}s, quadratic in time. In that case we can decompose the \textit{MSD}s into a purely diffusive part and a drift part.
\begin{equation}
MSD_i(\vec{r},t)=MSD_{i,Diff}(\vec{r},t)+MSD_{i,Drift}(\vec{r},t)
\label{eq:MSD_decomposition}
\end{equation}
where we can identify
\begin{equation}
MSD_{i,Drift}(\vec{r},t)=v_i^2(\vec{r})t^2.
\label{eq:Drift_contribution}
\end{equation}
All our previous considerations can then be applied to the purely diffusive part $MSD_{i,Diff}(\vec{r},t)$.
To determine the drift contribution and thus correct the \textit{MSD}s for the effect of drift, we need to measure the local velocities $v_i(\vec{r})$. This can be done by using the sampled displacements to calculate the local mean displacements (\textit{MD}s):
\begin{equation}
MD(\vec{r},t)\equiv \langle \Delta r_i \rangle(\vec{r},t)
\label{eq:MD_definition}
\end{equation}
For determining the local drift velocities from the \textit{MD}s, we need to account for the influence of systematic errors to the \textit{MD}s. Under the assumption that the tracking uncertainty is not biased, i.e. it has zero mean, it will not influence the \textit{MD}s. On the other hand we do need to take into account the finite image acquisition time $\theta$. Let us denote the actual particle positions with $\hat{\vec{r}}(t)$ as opposed to the measured positions $\vec{r}(t)$ which are subject to tracking errors.
Assuming the particle to move with constant velocity $\vec{v}$, the actual positions will be described by $\hat{r}_i(t)=\hat{r}_i(0)+v_it$ and the observed positions $r_i(t)$ for $t\geq \theta$ are therefore given by
\begin{align}
r_i(t)&=\frac{1}{\theta}\int_{0}^\theta \hat{r}_i(t-s)ds\\
&=v_i\left(t-\frac{\theta}{2}\right)+\hat{r}_i(0).
\label{shutter_influence_on_velocity}
\end{align}
The local velocities can thus be extracted by using the following relation:
\begin{equation}
v_i(\vec{r})=\frac{\langle \Delta r_i \rangle (\vec{r},t)}{t-\theta /2}
\label{eq:Driftvelocity}
\end{equation}
and their experimental error can easily be deduced from the standard error of the mean of the measured displacements. The calculated drift field for our experiment is shown in Figure~\ref{fig:flow_field_color_mapped}. In our case, flow was negligible within measurement accuracy $\left( \sqrt{\langle v_{x,y}^2 \rangle}=(0.5\pm0.6)~\mu m /s \right)$. The estimated drift contributions to the first two \textit{MSD} points remained below $4\%$ and was thus negligible compared to the undirected Brownian motion.


\section{Mapping the channel profile in 3D}
\begin{figure}[p]
		\includegraphics[width=8.5cm]{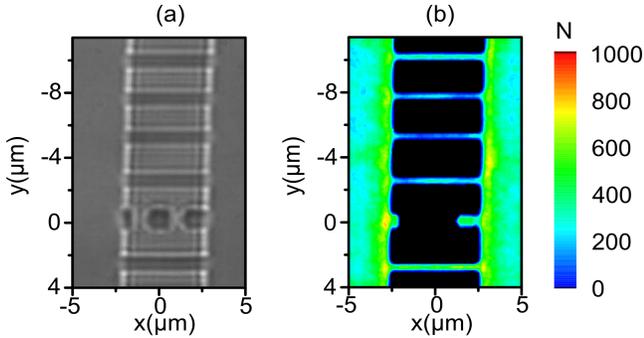}
	\caption{Mapping the system geometry in the plane by particle counting. (a) shows a bright-field image of the microfluidic channels. We mapped the space accessible to the particles by the following procedure. First, we divided the total area into position bins. Second, we tracked particles for 6~h of video, totaling 2.7 million recorded particle positions. Third, from all recorded particle trajectories each ´particle center position point was assigned to the corresponding position bin. Finally, we counted the number of these particle occurrences for each bin and assigned a color scale value to it. (b) presents this map for the region shown in (a). The map clearly visualizes the contours of the channels in the plane and shows a far stronger contrast than the bright-field image (a). The channels appear longer and thinner in (b) because it maps the region that is accessible to the particle centers and the finite particle radius separates this region from the PDMS walls.}
	\label{fig:countmaps}
\end{figure}

\begin{figure}[p]
	\centering
		\includegraphics[width=7cm]{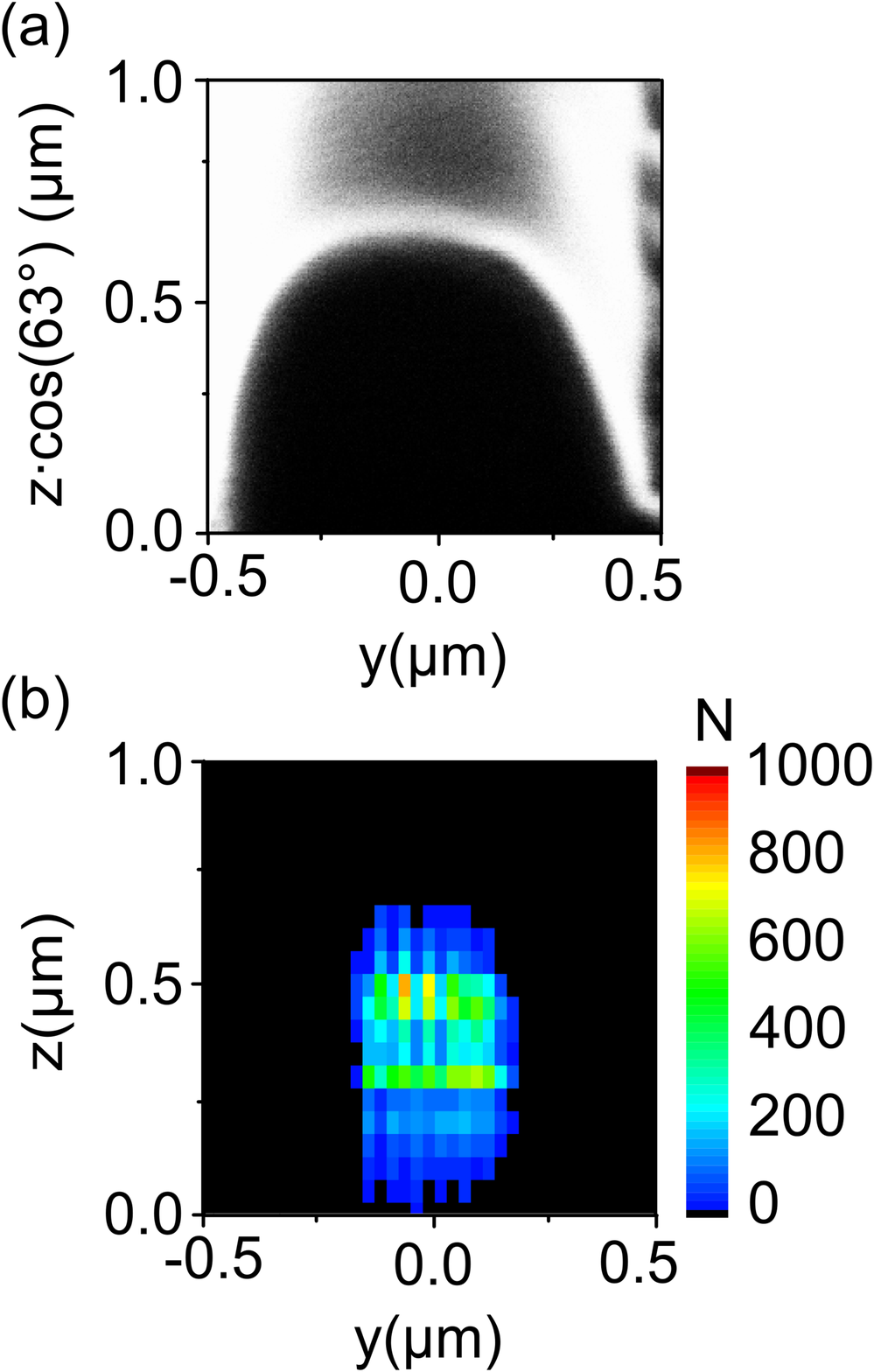}
	\caption{Mapping the channel cross-section by particle counting. (a) shows a scanning electron microscopy image of a typical Platinum wire which is used for creating the PDMS mold. The image was acquired at a tilt angle of 63$^\circ$ so that the scaling in $z$-direction is compressed by a factor of $cos(63^\circ)$. (b) The cross-section of the channel molded in PDMS from the Platinum wire (topmost channel in Figure~\ref{fig:countmaps}) is visualized by mapping the number of recorded positions by the same method as described in Figure~\ref{fig:countmaps}. Due to the limited tracking accuracy, counts also occur outside of the physically accessible region. We assume that these events are less frequent than tracked positions inside the physically accessible region (represented by the darker blue color in the map). By investigating the contour of higher counts (light blue, green, yellow and red) we recover the semi-elliptical cross-section of the Platinum wires. The bottom edge corresponds to the colloid touching the glass surface, i.e. the particle center at $z=0.25~\mu m$. Adding the finite particle radius to the region, we arrive at a channel cross-section of $800\times800~nm^2$.}
		\label{fig:crosssection}
\end{figure}

Determining the channel profile of the PDMS channels \textit{in situ} is a difficult task. The in-plane profile can be estimated from the bright-field images (Figure~\ref{fig:countmaps}(a)) but not least due to the poor optical contrast it is hard to accurately identify the channel edges. One reason for this weak contrast is that PDMS is largely transparent to visible light. Instead we propose a different technique to achieve \textit{in situ} mapping of the PDMS channel profile in 3D.\newline
In addition to calculating the local \textit{MSD}-vs-time we can assign all tracked particle positions to the corresponding bin. By counting the total number of assigned data points for each bin, we arrive at a position histogram $N(\vec{r})$ of the particle center positions. This provides a map of the system volume that is accessible to the particle centers. This can be done in the plane (Figure \ref{fig:countmaps}(b)) as well as in cross-section (Figure~\ref{fig:crosssection}(b)). The color scaled map for the plane (Figure~\ref{fig:countmaps}(b)) obviously provides a much stronger contrast than the bright-field image (Figure~\ref{fig:countmaps}(a)). The cross-sectional profile of the channels cannot be imaged directly via the bright-field images so the only option is to estimate the cross-section based on SEM images of the Platinum wires of the replica. However, since the PDMS structures relax after molding and bonding to the glass slide, their cross-section does not correspond exactly to that of the Platinum wires. So to get the cross-section of the PDMS channels itself it seems more reasonable to consider the position histograms as we have done in Figure~\ref{fig:crosssection}(b).

\section{Conclusions}
In this work we have presented methods for particle tracking and data analysis that allow simultaneous measurements of the spatial dependence of diffusion coefficients, tracking accuracies and drift velocities in the context of hindered diffusion in complex geometries. 
In our experiment we reported the measurement of local tracking accuracies in all three axial directions as well as the diffusion coefficients parallel to the channel axes. Significant flows were not observed. The importance of measuring these quantities locally rather than using global averaging methods is exemplified in Figures~\ref{fig:tracking_accuracies} and \ref{fig:Dx_colormap} where we see a strong spatial variation implying that a global average cannot be informative for any particular region of the system.
Generally, our methods enable fast and easy measurements of hindered diffusion coefficients in complex geometries without the need for specialized experimental equipment. 2D tracking requires solely a simple bright-field microscope with a CCD camera which should be available in most laboratories. The integration of a piezo stage is already sufficient to allow for full 3D resolution. As microfluidic techniques are well-developed, many different geometries can easily be created and the hindered diffusion coefficients measured within a few hours using the techniques presented here. Subsequently, the geometries found to exhibit an interesting spatial dependence of diffusion coefficients could be investigated at a higher resolution with single particle experiments using optical trapping and digital holography or total internal reflection microscopy as has been done for the simple geometry of spheres close to plane walls. While the particle tracking algorithms will differ from the one described in this work, our methods for statistical analysis of the trajectories will also be applicable to those experiments. On the other hand, the tracking algorithm can also be used for single particle tracking in fluorescence microscopy videos of motion that is not purely Brownian, such as colloids experiencing dielectrophoretic or electrokinetic forces. \newline
To conclude, we have provided the experimenter with essential tools for single particle tracking studies of phenomena such as hindered Brownian motion in complex geometries or dielectrophoresis. We hope that these tools will help to build a better understanding of biological transport processes and be useful in technical applications such as the dielectrophoretic separation and characterization of micro-organisms.

\begin{acknowledgments}
S.L.D. acknowledges funding from the German Academic Exchange Service~(DAAD) and the German National Academic Foundation.  S.P. and U.F.K. were supported by an ERC starting grant. S.P. also acknowledges the support from the Leverhulme Trust and the Newton Trust through an Early Career Fellowship. 
\end{acknowledgments}

\FloatBarrier
%

\end{document}